\documentclass[twocolumn,showpacs,preprintnumbers,aps,pra,floatfix,10pt,superscriptaddress]{revtex4-1}
\pdfoutput=1 
\usepackage{subfigure}
\usepackage{graphicx}
\usepackage{color}
\usepackage{epsfig}
\usepackage{amsmath}
\usepackage{amssymb}
\usepackage{verbatim}
\usepackage{graphics}
\usepackage{natbib}

\usepackage{float}
\floatstyle{ruled}

\usepackage{pstricks}
\usepackage{eepic}

\begin{document}

\title{Critical-like Features of Stress Response in Frictional Packings}

\author{Abdullah Cakir} 
 \email{acakir@ntu.edu.sg}
\affiliation{Division of Physics and Applied Physics, School of Physical and
  Mathematical Sciences, Nanyang Technological University, Singapore}

\author{Leonardo E. Silbert}

\affiliation{Department of Physics, Southern Illinois University Carbondale,
  Carbondale, Illinois 62901, USA}

\begin{abstract}
  The mechanical response of static, unconfined, overcompressed face centred
  cubic, granular arrays is studied using large-scale, discrete element method
  simulations. Specifically, the stress response due to the application of a
  localised force perturbation - the Green function technique - is obtained in
  granular packings generated over several orders of magnitude in both the
  particle friction coefficient and the applied forcing. We observe crossover
  behaviour in the mechanical state of the system characterised by the
  changing nature of the resulting stress response. The transition between
  anisotropic and isotropic stress response exhibits critical-like features
  through the identification of a diverging length scale that distinguishes
  the spatial extent of anisotropic regions from those that display isotropic
  behaviour. A multidimensional phase diagram is constructed that parameterises
  the response of the system due to changing friction and force perturbations.
\end{abstract}
\maketitle
\section{Introduction}
Granular materials are ubiquitous throughout industry and nature and
constitute myriad materials ranging from snowpacks, planetary regoliths, silos
filled with grain, through to pharmaceutical pills and ceramic components. Yet
predicting the mechanical properties of a given granular material, such as a
packing of frictional grains, is difficult due to the many factors that
influence the stability of the packing. These can include microscopic features
such as particle roughness, shape, and softness, to more macroscopic issues
like the overall structural arrangement of the grains, density or packing
fraction of the granular medium, and the properties of the container holding
the material. Indeed, through a number of experimental
\cite{rajchenbach2,Geng2003274,10.1007/s101890170019,nagel6,nagel7,PhysRevLett.86.3308,PhysRevE.83.021304}
and theoretical and computer simulation studies
\cite{PhysRevE.67.031302,Nature,10.1140/epje/i2006-10012-6,socolar1}, it has
emerged that a granular medium can appear to behave not only as a mechanically
stable continuous elastic solid but can also exhibit strongly non-linear
features characterised by highly heterogeneous stress properties. Though more
recently, simulations that take into account the full preparation history of
pouring a granular piling under gravity find that traditional elastoplastic
continuum models are able to reproduce the stress-dip phenomena
\cite{10.1103/PhysRevLett.113.068001}. However, in the absence of such
knowledge it is unclear how best to categorise any given granular
packing. Yet, despite these complications, there are several methods that can
be used to probe the mechanical robustness of the material.

\begin{figure}[htbp]
  \begin{center} 
    \includegraphics[scale=0.275]{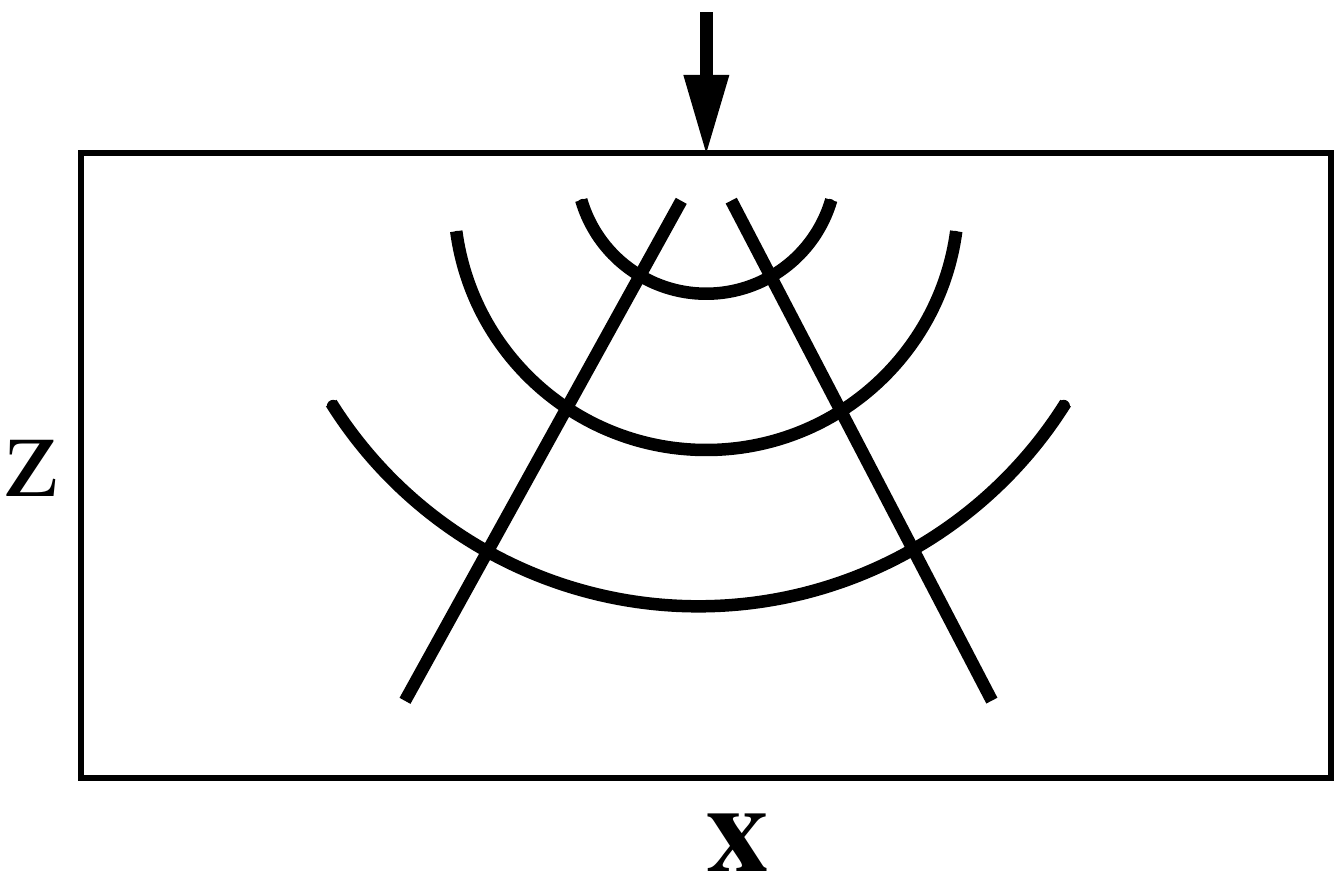}
    \caption{\small Implementing the force perturbation protocol referred to
      as the Green function technique. In this illustration, the arrow
      represents a downward, vertical point force applied at the middle on the
      top of a slab of material under interrogation. Possible stress response
      profiles are represented by the arcs and directed lines.  The mechanical
      state of the systems is then inferred through monitoring the resulting
      stress profile along the $x$-direction at different slices in $z$.
      Note, this is merely a 2d visualisation of the 3d problem we discuss
      here.}
    \label{fig0}
  \end{center}
\end{figure}

One protocol readily implemented that provides just such information on the
nature of the stress state of a granular medium is the Green function
technique: apply a localised, point force to the packing and observe the
resulting stress profile in response to the perturbation \cite{timoshenko1}. A
schematic of the procedure is shown in Figure \ref{fig0}. This technique has
been implemented in experiments \cite{Geng2003274} and computer simulations
\cite{springerlink:10.1140/epje/i2005-10002-2,Nature,PhysRevE.77.041303} to
show that two dimensional (2d) disc packings tend to agree with traditional,
isotropic, elasticity theory \cite{Johnson,timoshenko1,Landau} when particle
friction is large. That is to say, the maximum in the measured stress profile
remains directly `below' the location of the perturbing force and is often
termed ``one-peak'' response in reference to two dimensional studies. This
becomes a single lobe in three dimensions (3d). However, for smoother
particles, the stress response becomes highly anisotropic, exhibiting a
``twin-peak'' (ring profile in three dimensional packings) of maximum stress
that is not directly in line with the applied force. Nevertheless, despite
these differences, either type of response is generally regarded to belong to
solutions of the elastic-elliptic class
\cite{PhysRevE.67.031302,PhysRevE.77.041303,mliu3}. Furthermore, some studies
\cite{Geng2003274,Nature} have also hinted at the possibility that the
response within a grain pile may change character as one looks further from
the location of the perturbing source.
In another study \cite{PhysRevLett.97.208001.2006}, a transition in the stress
response within granular arrays is associated with a length scale though there
has not been an explicit identification of the length scale nor the dependence
on friction coefficient. This has important consequences as to whether a given
material will indeed conform to a more isotropic or anisotropic picture, and
raises the possibility that the stress state of a particulate medium may be
tailored to respond differently on varying length scales.  As yet there
remains no method to predict \emph{a priori} what the expected stress response
of a three dimensional granular packing is and how the stress state of the
system varies within the packing itself.

It is precisely this issue that we address here. In an effort to reduce the
large parameter space of all possible influences, we focus on what we reason
are the two most dominant effects: particle friction and magnitude of the
perturbing force. (The role of structural disorder was addressed previously
\cite{10.1007/s10035-009-0156-0}.) This study is needed to provide a guide in
the design of granular solids where they can behave as a stable solid or
exhibit strongly non-linear features with tailored mechanical properties. From
this study, length scales distinguishing between anisotropic from isotropic
behaviours can be estimated. The organisation of this paper is as follows. The
Boussinesq equations are briefly discussed in section~\ref{boussi}. The manner
in which we implement the packing preparation method and the Green function
technique are described in section~\ref{simu} and summarised in
Ref.~\cite{1742-5468-2011-08-P08005}. The results of this paper are presented
in section~\ref{results} followed by closing discussions and conclusions,
section~\ref{concl}.

\section{Boussinesq Equations}
\label{boussi}
Isotropic linear elastic theory is based on the solutions of the elliptic
differential equations of mechanical equilibrium \cite{Landau}.  For a point
force ($F_{\rm app}$) applied in downward z direction to an elastic half space
- the Green function technique (see Figure \ref{fig0}) - the solutions of the
elliptic mechanical equilibrium differential equations provide the
displacement field
\begin{equation}
  \begin{tabular}{ccl}
    $u_{x}$ & $=$ & $\frac{F_{\rm app}}{4\pi G}\left( \frac{xz}{r^{3}} -
      (1-2\nu)\frac{x}{r(r+z)}\right)$, \\ \\
    $u_{y}$ & $=$ & $\frac{F_{\rm app}}{4\pi G}\left(\frac{yz}{r^{3}} -
      (1-2\nu)\frac{y}{r(r+z)}\right)$, \\ \\
    $u_{z}$ & $=$ & $\frac{F_{\rm app}}{4\pi G}\left(\frac{z^{2}}{r^{3}} + 
      \frac{2(1-\nu)}{r}\right)$,
  \end{tabular}
  \label{equat1}
\end{equation}
where, $u_{x,y,z}$ are the $x$, $y$, and $z$ displacement field components and
$r = \sqrt{x^2+y^2+z^2}$. $\nu$ and $G$ are bulk material parameters, the
Poisson ratio and modulus of rigidity, respectively. $\nu$ is restricted to
the values between $-1$ and $0.5$ \cite{Landau}.  The above equations,
Eqs.~\ref{equat1}, express a displacement vector at an arbitrarily chosen
point inside a solid in response to the localised force perturbation
$F_{app}$.

By implementing the displacement field components into the definitions of
strain and applying Hooke's law between stress and strain, we arrive at the
stress response components known as Boussinesq equations (in three dimensions)
\cite{Landau,Johnson}. The origin of the coordinate system is at the
application point of the applied forcing. Since the force is in the z
direction the stress component characterising the response of a system is
$\sigma_{zz}$, which we refer to as $\sigma$,
\begin{equation}
  \sigma = \frac{3F_{\rm app}}{2\pi}\frac{z^3}{r^{5}}
  \label{seq2}
\end{equation}
The crucial feature of the Green function approach is in reducing the problem
to a material parameter free equation, Eq.~\ref{seq2}. The procedure we follow
is through direct comparison of our simulation data with those of the
Boussinesq result, Eq.~\ref{seq2}. When we have ``single-peak'' behaviour (in
three dimensions, a lobed profile) we attempt to fit our results. We repeat
this process over our range of parameters in friction coefficients and force
perturbations.

\section{Simulation Procedure}
\label{simu}
\subsection{Packing Generation Protocol}
We employ simulation domains with periodic boundary conditions to reduce the
effects of confining walls which can strongly influence the nature of the
stress response in small systems \cite{1742-5468-2011-08-P08005}. From our
previous studies on confined systems, we found that for a granular packing
that exhibits an isotropic, elastic-like response, the response behaviour
converged to the theoretical predictions for systems of linear dimension, $L
>20 d$, for particle diameter $d$. Therefore, for this study we generated
particle configurations of size $L \approx 60d$, as a system that is suitable
to mimic the response within a semi-infinite medium. We placed $N=256000$,
monodisperse, inelastic spheres at the lattice points of a face centred cubic
(FCC) array in the absence of gravity. The packing was then isotropically
compressed under uniform pressure until the system reached a packing fraction,
$\phi = 0.742$, which is slightly above the hard-sphere FCC value,
$\phi_{FCC}= \pi/\sqrt{18}$. Particles interact only on contact through a
linear force law characterised by stiffness parameters, $k_{n}$ and $k_{t}$,
for interactions normal and tangential to the contact plane where the maximum
friction force is determined by the particle friction coefficient $\mu$. For
this study, the particle Poisson ratio was set to, $k_{t}/k_{n} = 1$. Friction
is included using a standard granular dynamics routine \cite{SilbertBook}.
The normal and tangential contact forces between particles ($F_{n,t}$) are
given by,
\begin{equation}
  \begin{tabular}{ccc}
    $F_{n}$ & = & $-k_{n}\Delta_{ij} - m_{\rm eff} \gamma^{n}v_{n_{ ij}}$\\ \\
    $F_{t}$ & = & $-k_{t}S_{ij}- m_{\rm eff} \gamma^{t}v_{t_{ij}}$.
    \label{forces}
  \end{tabular}
\end{equation}
where $\Delta_{ij} \equiv (r_{ij} - d)$ is the surface compression of two
contacting particles and $v_{{n,t}_{ij}}$ are the relative normal and
tangential velocities of two particles. $r_{ij}$ is the relative distance
between particles $i$ at position $r_{i}$ and $j$ at position $r_{j}$.  The
effective mass, $m_{\rm eff}$ is equal to $m/2$, for particle mass
$m$. $\gamma^{n,t}$ are the normal and tangential damping factors, which
together with the $k_{n,t}$ and $m$ set the coefficient of restitution of the
particles, which here is fixed to $\approx 0$ \footnote{We used a large
  inelasticity to dissipate the kinetic energy from the packing and return the
  system to a mechanically stable as quickly as possible after the
  perturbation.}. $S_{ij}$ represents the integrated displacement of the
contact point of the particles $i$ and $j$ while the two particles remain in
contact. The Coulomb condition is written as $F_{t} \leq \mu F_{n}$.  We
generated static packings for different friction, ranging over several orders
of magnitude, $\mu = 0, 0.001, 0.002, 0.005, 0.01, 0.02, 0.05, 0.1, 0.2, 0.5,
1, 10$. Below, we present our results in simulation units where mass, length,
and force values are appropriately scaled by the particle properties: mass $m
= 1$, diameter $d = 1$, and stiffness $k = k_{n} =k_{t}= 1$.

\subsection{Force Perturbation Protocol}
\label{forcepert}
\begin{figure}[htbp]
  \begin{center} 
(a)
\includegraphics[scale=0.22]{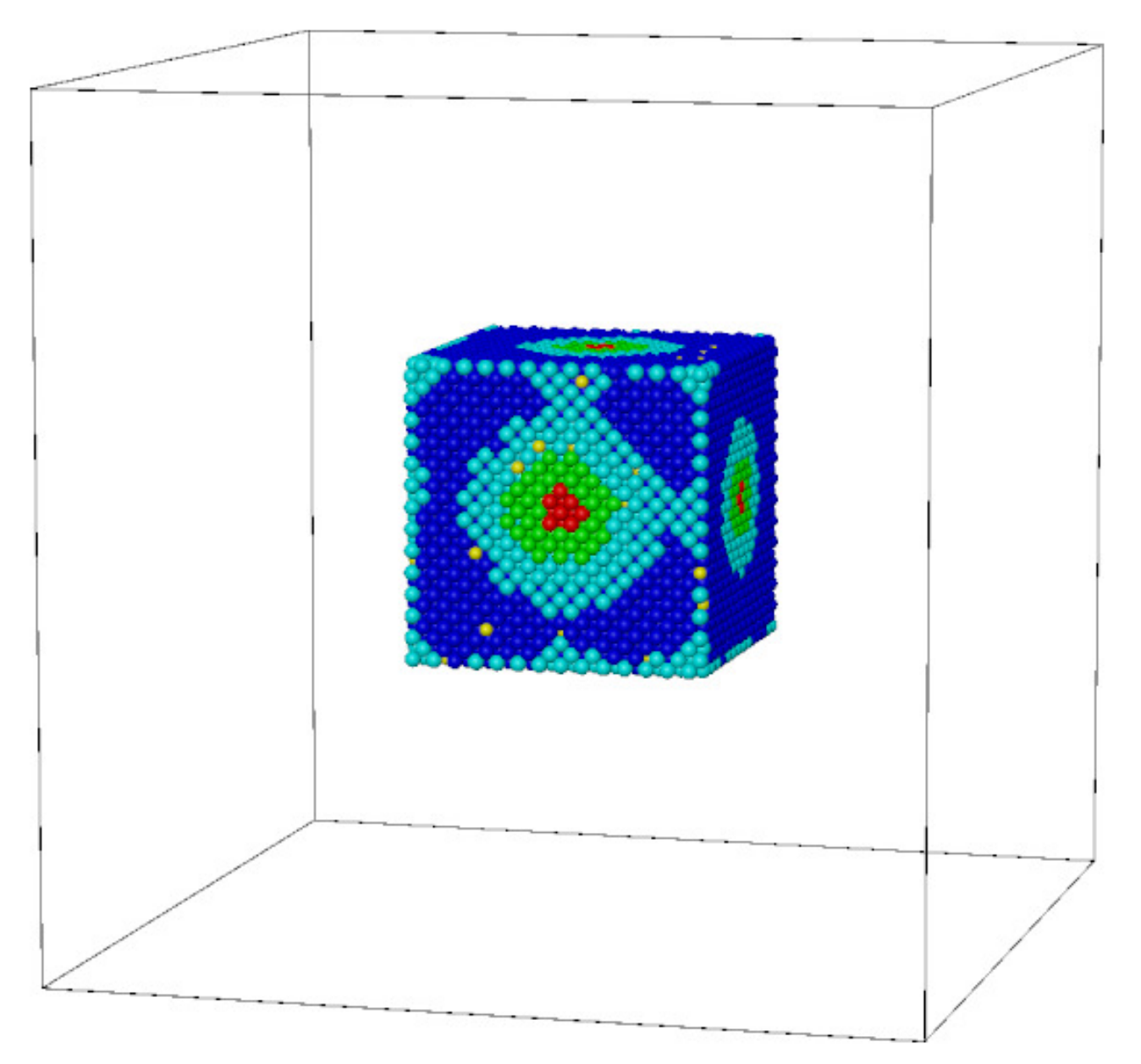}
(b) 
\includegraphics[scale=0.22]{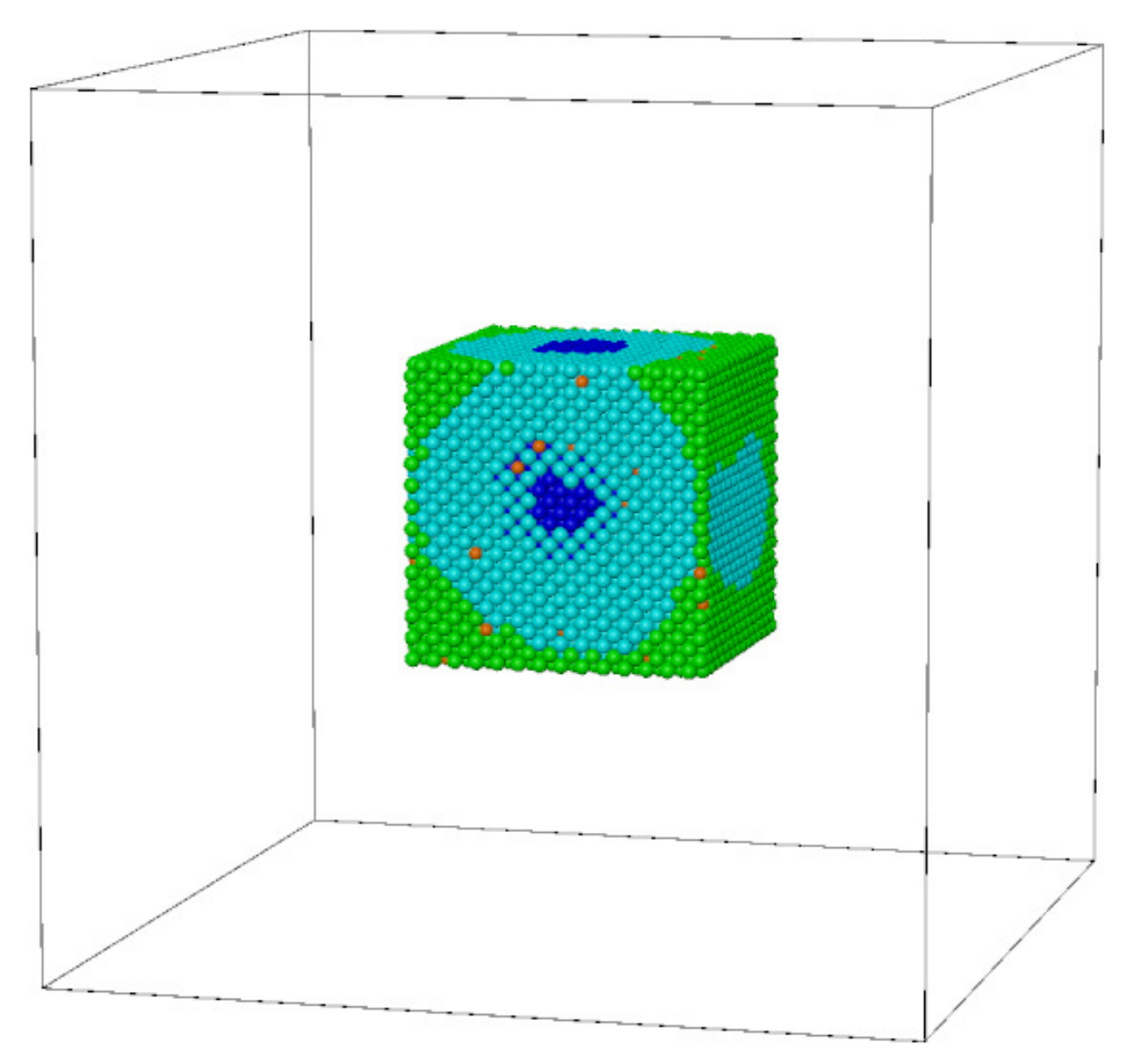}
\caption{Simulation snapshots of the force perturbation in systems that
  exhibit, (a) an anisotropic, ring-like (twin-peak) response $\mu=0$ and (b)
  an isotropic, single lobe (single-peak) response, $\mu=1$. Each panel
  focuses on the central region of the packings with particles shaded by their
  displacement relative to the unperturbed initial state. A force perturbation
  was applied at the centre of the box as described in the text. Shading
  represents magnitude of the displacement (relative to the average for that
  system): blue end of spectrum corresponds to larger displacements. For the
  FCC particle arrangement at an overcompressed packing fraction $\phi=0.742$
  used in this study.}
    \label{sfig1}
  \end{center}
\end{figure}
The localised force perturbations were implemented in the following manner. We
first located an approximately spherical cluster of particles within the
central region of the packing that typically consisted of $<0.3\%$ of the
total number of particles in the system. We then increased their diameters by
a specified amount and allowed the packing to relax back into a mechanically
stable state keeping the same periodic boundary conditions. The effect of this
local swelling of particles is equivalent to applying a localised and
isotropic perturbing force, $F_{\rm app}$, at the centre of the packing, that
we varied over several orders in magnitude: $F_{\rm app}= 0.0005, 0.001,
0.0025, 0.005, 0.01, 0.025, 0.05, 0.1, 0.25~kd$. This procedure, initially
implemented in Ref.~\onlinecite{10.1007/s10035-009-0156-0}, was motivated by
several considerations. Firstly, given that we exclude gravity in this work,
we preferred to avoid imposing a uni-directional force that would break the
symmetry of our set up. We also reasoned that this protocol offers a practical
means to implement force perturbations in real systems. As a case in point,
during the course of this work, a recent experimental work has implemented a
similar procedure in studies of the mechanical properties of two-dimensional
granular packings \cite{10.1103/PhysRevLett.113.198001}. We have verified that
the method of increasing the size of particles, located within a small region
centred at origin, does indeed result in a corresponding force magnitude equal
to $F_{\rm app}$, by measuring the average force over a spherical shell in the
vicinity of the inflated region. Representative snapshots of how the force
perturbation influences the packing are shown in Figure~\ref{sfig1}. These
images display the particles occupying a sample, cubic central portion of the
packing (out to 10d) for $\mu = 0$ and $F_{\rm app} = 0.25$
(Figure \ref{sfig1}(a)) and $\mu = 1$ and $F_{\rm app} = 0.02$
(Figure \ref{sfig1}(b)). These examples were chosen to clearly illustrate the
change in response. The magnitude of particle displacements relative to their
initial, unperturbed positions are indicated by a shading scale, with the blue
end of the spectrum indicating larger displacements. These microdisplacements
of the particles in response to the applied force indicate the reasons behind
the differences between the anisotropic and isotropic response functions. In
the anisotropic case, maximal particle displacements are directed along an
open cone, whereas in the isotropic case the displacements ultimately form a
closed `ball' emanating away from the region of the perturbing source. We have
verified that for the system size studied here, our results match exactly the
predictions of Eq.~\ref{seq2} when we observe an isotropic response (which
occurs for large friction and small applied force - see Figure \ref{sfig4}).

We define the stress response as the difference between the final and initial
stress states.  We computed the stress in the packing before, $\sigma_{\rm
  i}$, and after, $\sigma_{\rm f}$, the imposition of the perturbation, so
that the resulting stress response is,
\begin{equation}
  \sigma = \sigma_{\rm f} - \sigma_{\rm i}. 
\label{stresseq}
\end{equation}

\subsection{Coarse Graining Procedure}
To be able to accurately compare our simulation stress response with
elasticity theories, we implement a stress computation method that connects
our microscopic, discretized calculations with macroscopic, continuum length
scales. A stress expression with a coarse-graining procedure developed by
Goldhirsch and colleagues \cite{10.1007/s10035-010-0181-z,PhysRevE.77.041303}
gives continuum stress components. We chose a Gaussian coarse-graining
function, so that the components $\alpha,\beta$ (=$x,y,z$), of the stress
tensor $\sigma({\bf r})$, are obtained from,
\begin{equation}
  \sigma_{\alpha \beta}({\bf{r}})=\frac{1}{2}\sum \limits_{i} \sum \limits_{j \neq i} F_{ij \alpha}
  r_{ij \beta} \int \limits_{0}^{1} \frac{ds}{ (\pi^{1/2} \omega)^{3}} e^{-(\mid
    {\bf r}-{{\bf r}_{i}} + s {{\bf r}_{ij}} \mid /\omega)^{2}}
  \label{seq5} 
\end{equation} 
where $\omega$ is the coarse-graining length scale and ${\bf F}_{ij}$ are the
contact forces between particles $i$ and $j$. This expression is normalised.
The value of the coarse-graining length scale, $\omega$, is important in terms
of obtaining a fine resolution and yet continuum description of stress
components. Here, the particle diameter, $d$, is a suitable choice. The coarse
grained stress expression, Eq.~\ref{seq5}, is calculated on a spatial grid
discretized by grid size of $0.2d$.
\section{Results}
\label{results}
\subsection{Stress Maps}
\begin{figure*}[htbp]
\onecolumngrid
\begin{center}   
  \includegraphics[scale=0.55]{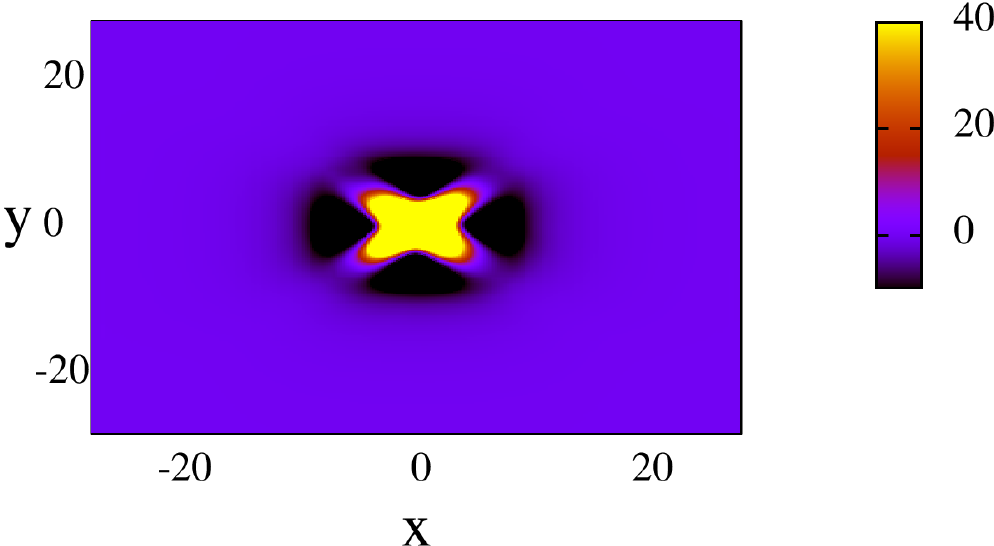}
  \includegraphics[scale=0.55]{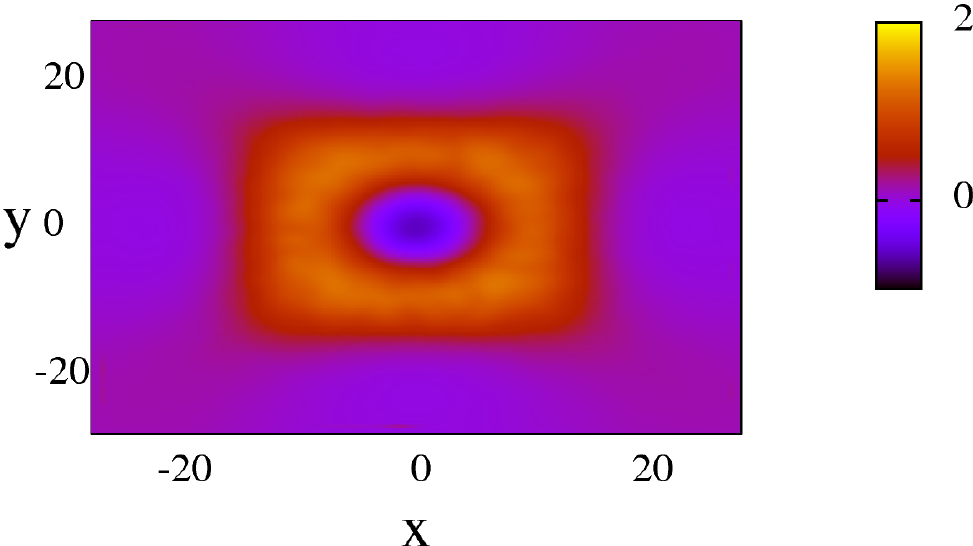}
  \includegraphics[scale=0.55]{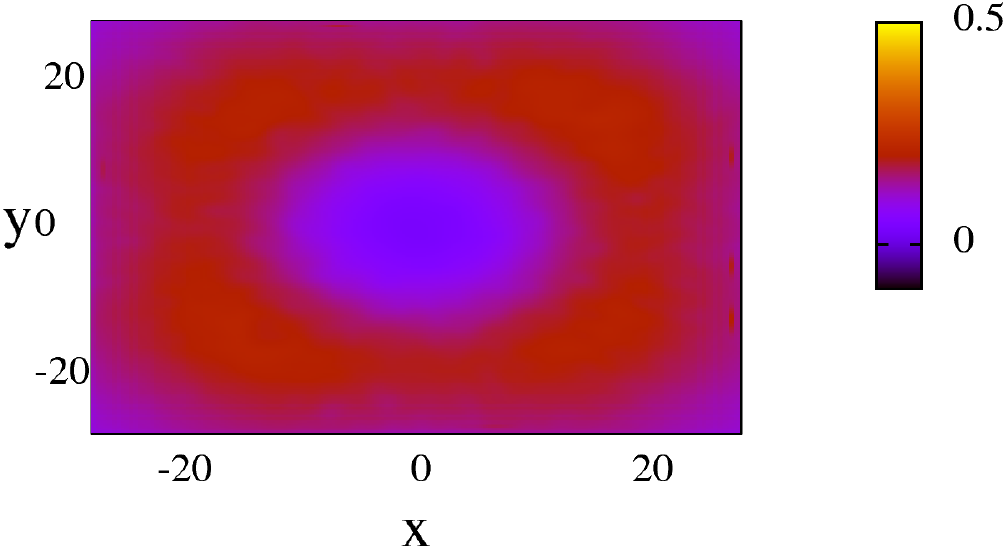}
  \includegraphics[scale=0.55]{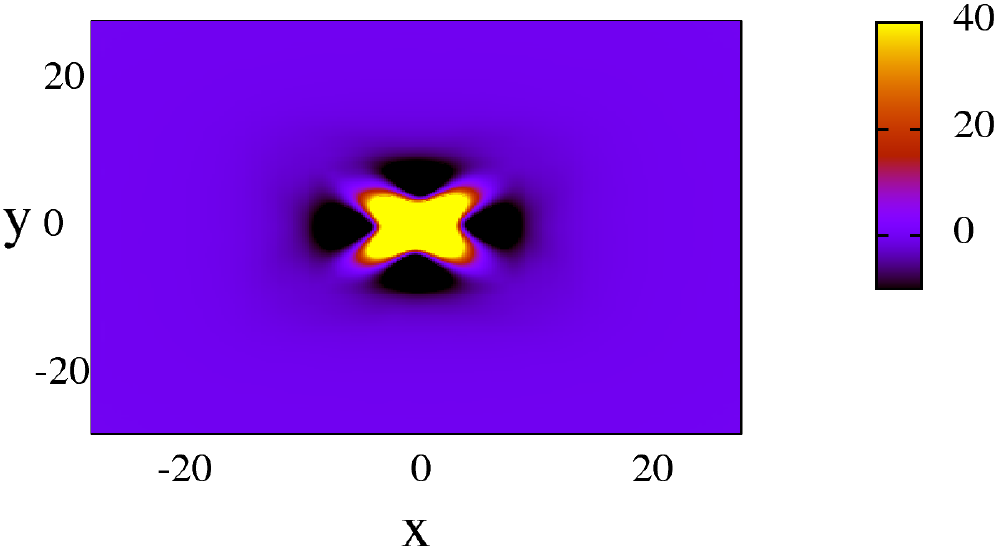}
  \includegraphics[scale=0.55]{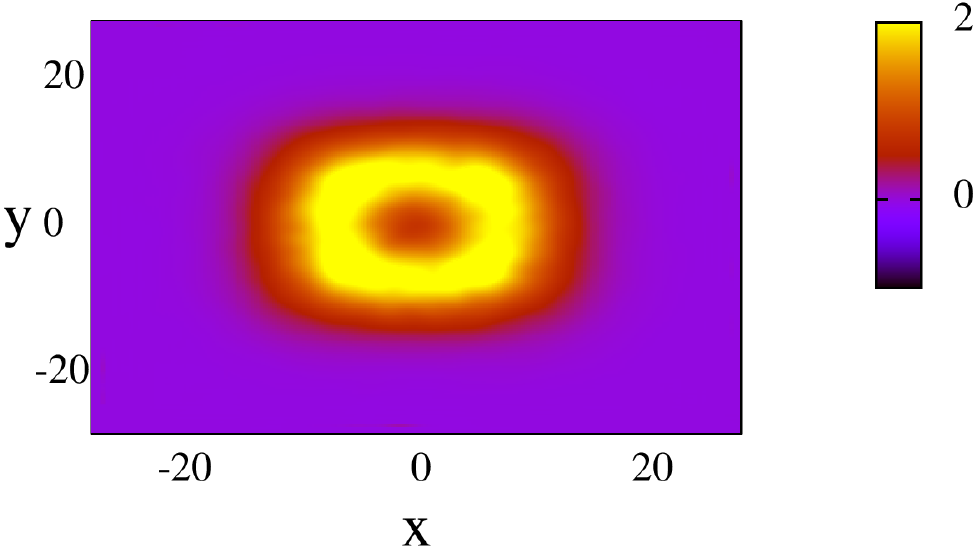}
  \includegraphics[scale=0.55]{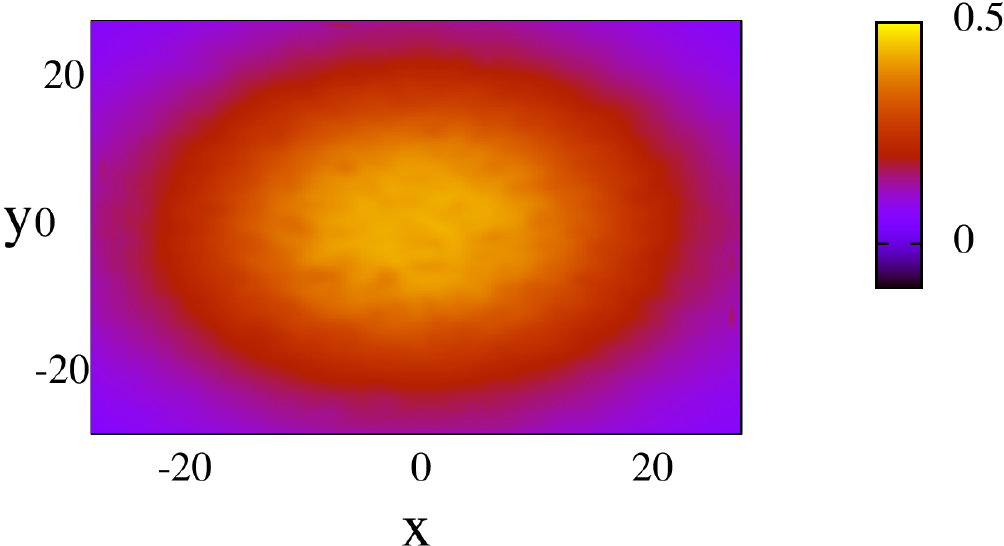}
  \includegraphics[scale=0.55]{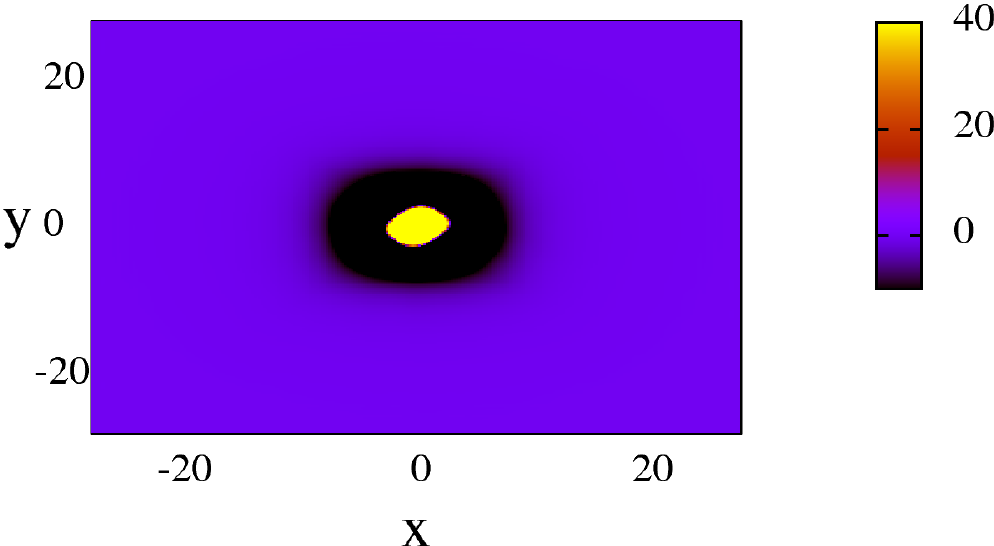}
  \includegraphics[scale=0.55]{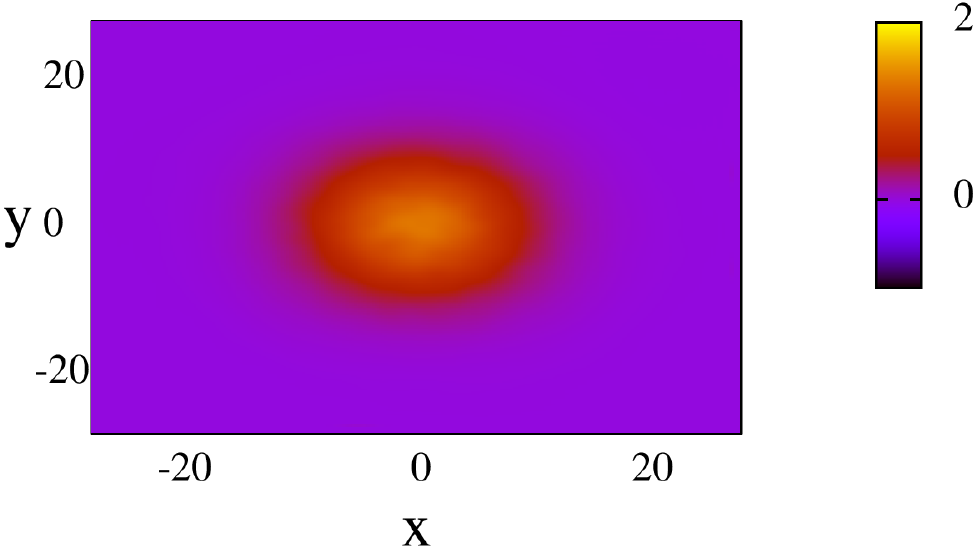}
  \includegraphics[scale=0.55]{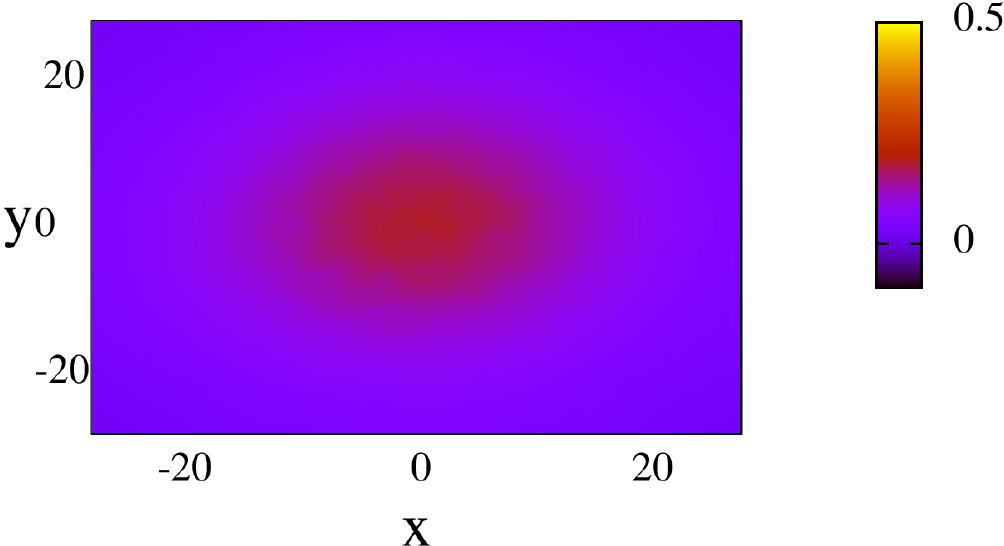}
  \caption{\small The stress response viewed in planes normal to the direction
    of the perturbing force. The origin of the coordinate system is defined as
    the location of the perturbing force. Here, the horizontal slices are
    denoted $xy$-planes and the different columns correspond to different
    slices in the $z$ direction. The three rows correspond to different
    particle friction: $\mu=0$ (top), $\mu=0.05$ (middle), and $\mu=10$
    (bottom). Within each row, columns show the stress at different distances
    from the location of the perturbing force: left column is closest to the
    force. The shading scale is the same for each row but different among the
    columns to better reveal the stress profiles. In all cases, the applied
    force magnitude, $F_{app}=0.25 kd$.}
  \label{fig1}
\end{center}
\twocolumngrid
\end{figure*}
To make our results connect visually with existing experiments, we present our
results as stress maps in different planes with respect to the perturbation
source. We also plot the circularly-averaged stress as two dimensional
profiles across the central section from planes that are perpendicular to the
direction of the force perturbation. In Figure~\ref{fig1} we show stress
profiles for packings with different friction subject to an applied force
perturbation of a relatively large, fixed magnitude $F_{\rm app} = 0.25~kd$.
The panels of Figure~\ref{fig1} are arranged as follows: each row corresponds
to a single friction coefficient while each column shows data at different
distances from the perturbing source.  We refer to these planes as `top',
`middle', and `bottom' referring to how far from the perturbation source we
view the stress, with the top layer closest to the source and bottom furthest
away. The left column is stress profile closest to the source (`top') and the
right column furthest away (`bottom').  Immediately we see how friction plays
a major role in determining the response of the system. Consistent with
previous studies \cite{Nature} we find that friction enhances the elastic
nature of the response whereby the response profile remains ``single peaked''
at all distances from the source, with linear, isotropic elastic broadening of
the profile. At zero friction, the profile is strongly anisotropic away from
the source with a ring-like or ``twin-peak'' response, reflecting the lattice
structure of the underlying packing, that persists at the largest
distances. However, for intermediate friction values, the response initially
resembles the zero-friction state out to a finite distance from the source
then transitions to a more isotropic, single-peak response at larger
distances.
\vspace{-0. in}
\subsection{Stress Profiles} 
\vspace{-0. in}
\begin{figure}[htbp]
  \begin{center} 
    \includegraphics[scale=0.5]{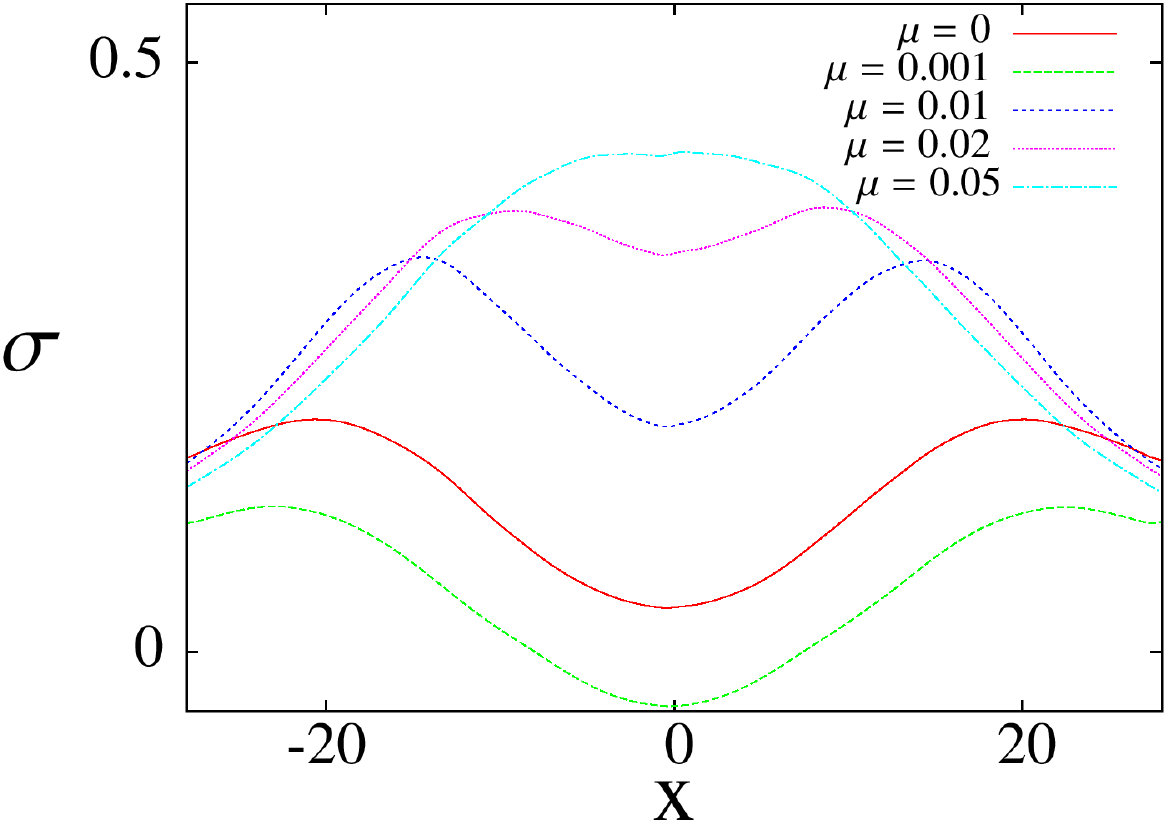}
    \caption{Circularly averaged stress profiles $\sigma$, at the `bottom'
      layer due to applied force $F_{\rm app} = 0.25 kd$, for different
      friction coefficient $\mu$, as indicated in key.}
    \label{sfig2}
  \end{center}
\end{figure}
\begin{figure}[htbp]
  \begin{center} 
    \includegraphics[scale=0.5]{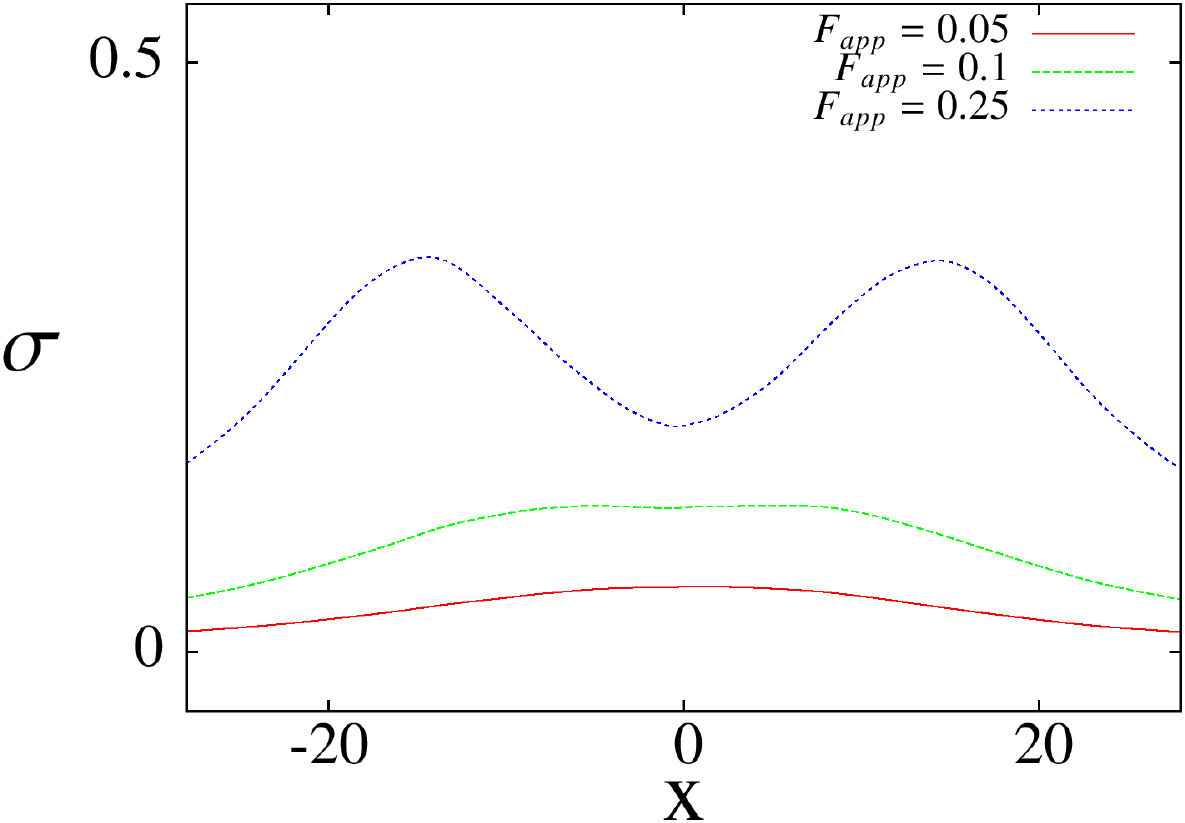}
    \caption{Circularly averaged stress profiles $\sigma$, at the bottom layer
      due to forces, at $\mu=0.01$, for different applied force $F_{app}$.}
    \label{sfig3}
  \end{center}
\end{figure}
\begin{figure*}[thbp]
\onecolumngrid
\begin{center}
{\bf a}
\includegraphics[scale=0.55]{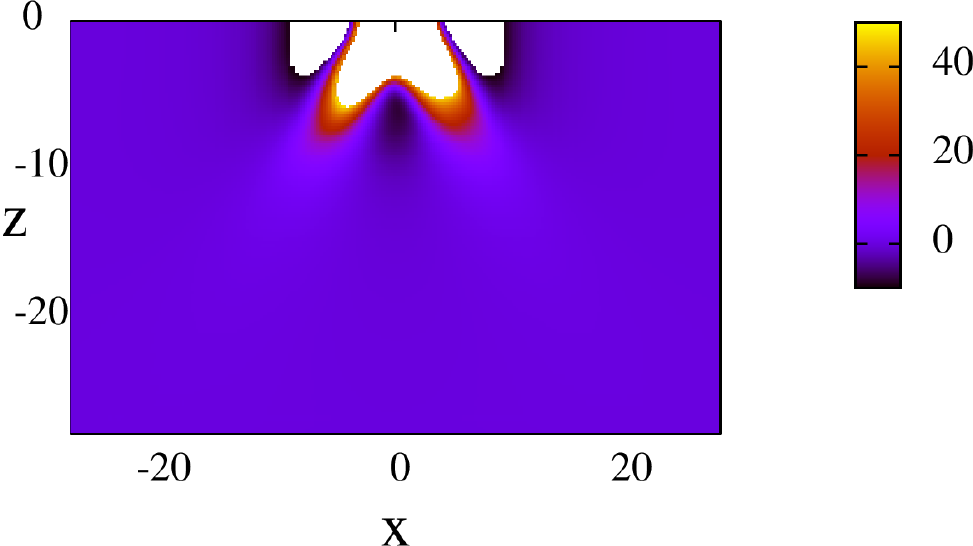}
{\bf b}
\includegraphics[scale=0.55]{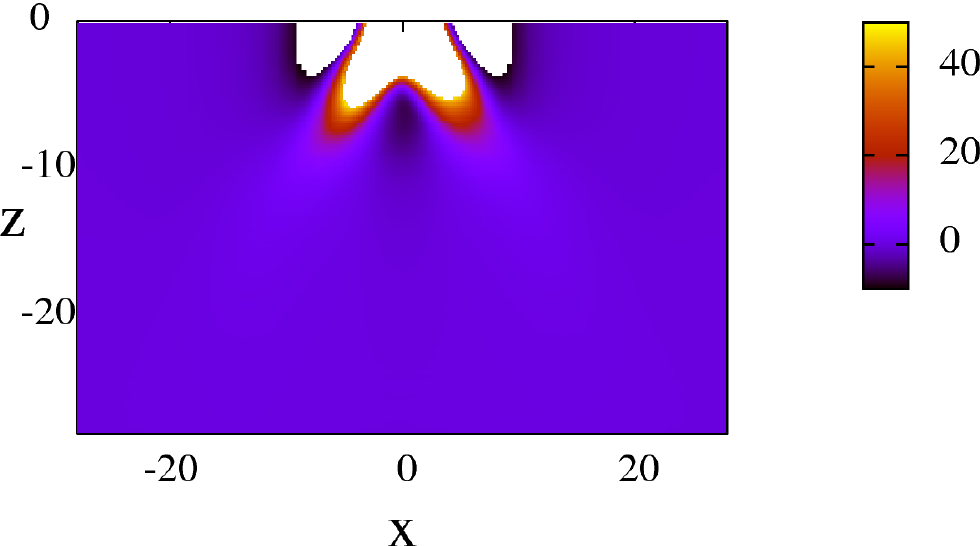}
{\bf c}
\includegraphics[scale=0.55]{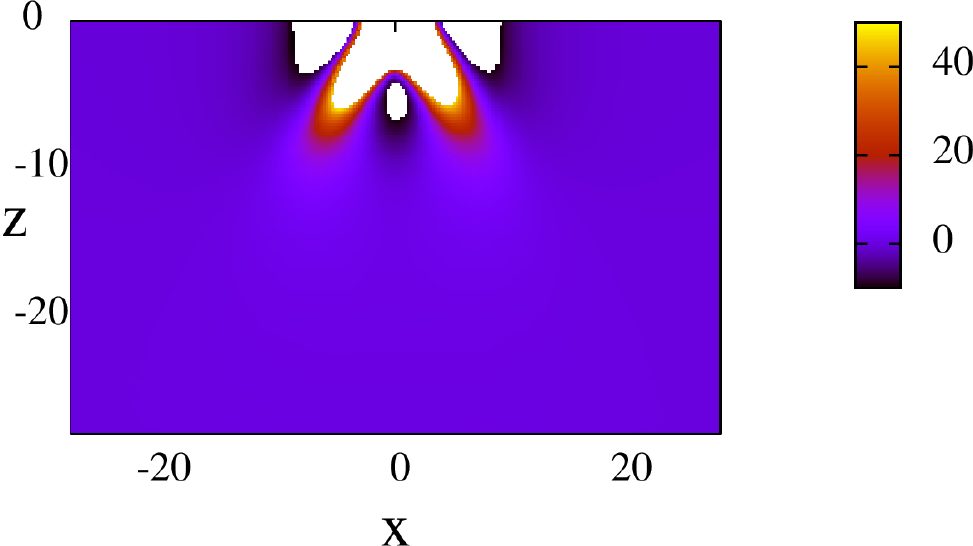}\\
{\bf d}
\includegraphics[scale=0.55]{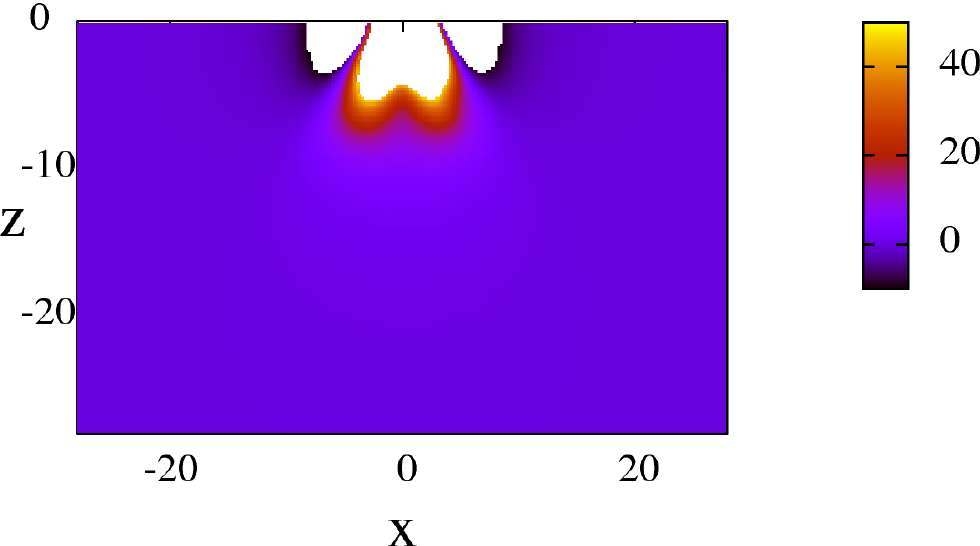}
{\bf e}
\includegraphics[scale=0.55]{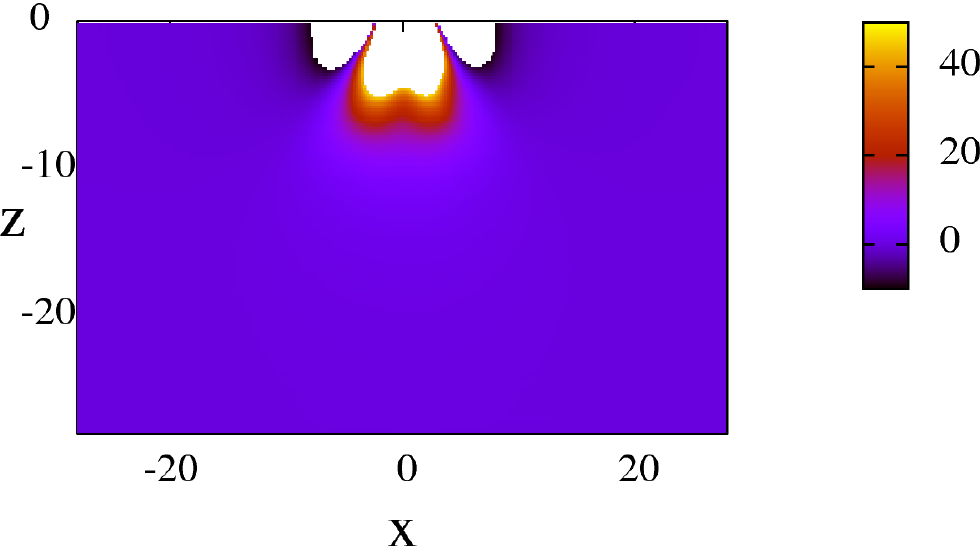}
{\bf f}
\includegraphics[scale=0.55]{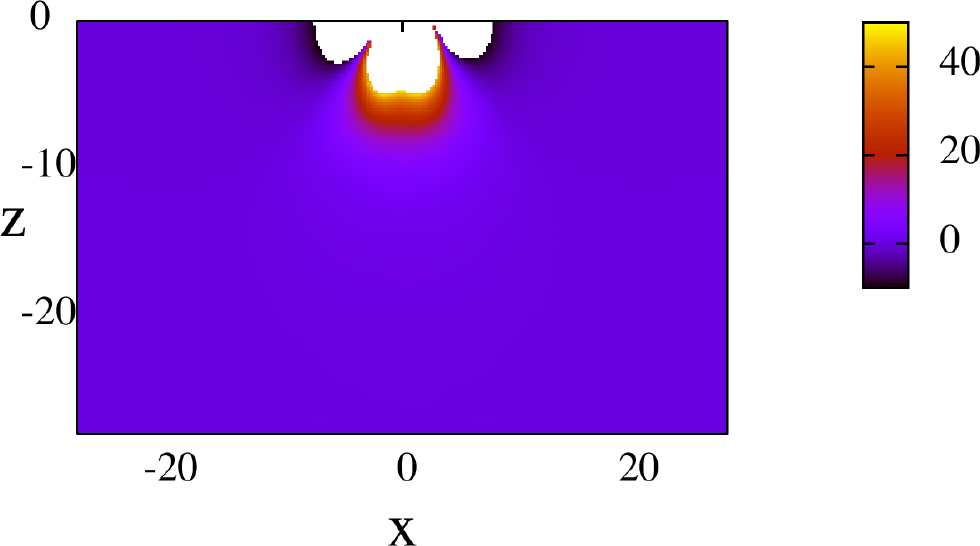}
\caption{\small The influence of friction on the stress response. Profiles are
  viewed in planes parallel to the direction of the applied force
  perturbation, for a force magnitude $F_{app}=0.25 kd$. Panels are particle
  friction values: (top left) $\mu=0$, (top middle) $\mu=0.01$, (top right)
  $\mu=0.05$, (bottom left) $\mu=0.5$, (bottom middle) $\mu=1$, and (bottom
  right) $\mu=10$.}
\label{fig2}
\end{center}
\twocolumngrid
\end{figure*}
\begin{figure}[htbp]
  \begin{center} 
    \includegraphics[scale=0.5]{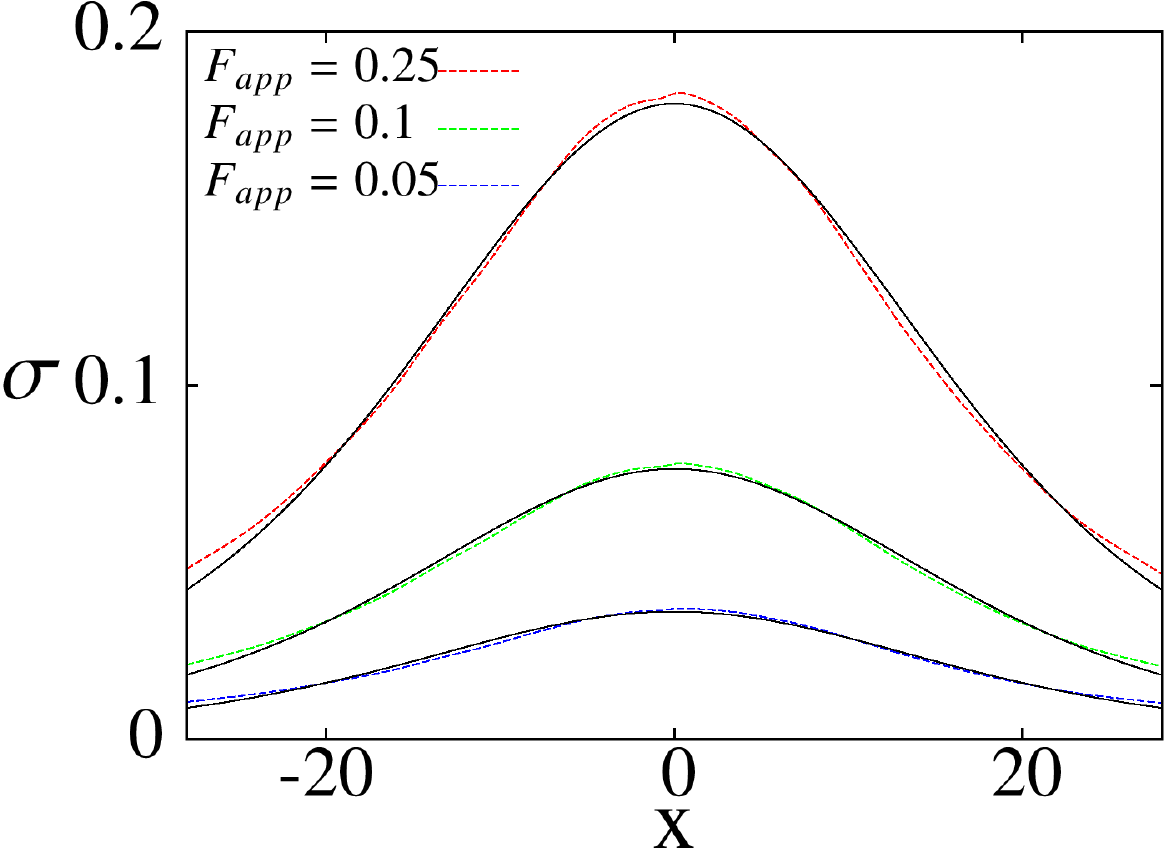}
    \caption{Circularly averaged stress profiles $\sigma$, at the bottom layer
      due to forces, with $\mu=10$, for different applied force $F_{app}$ (color
      dashed lines) and the fittings to the Boussinesq equation (solid black
      lines).}
    \label{newfg}
  \end{center}
\end{figure}
\begin{figure}[htbp]
  \begin{center} 
    \includegraphics[scale=0.5]{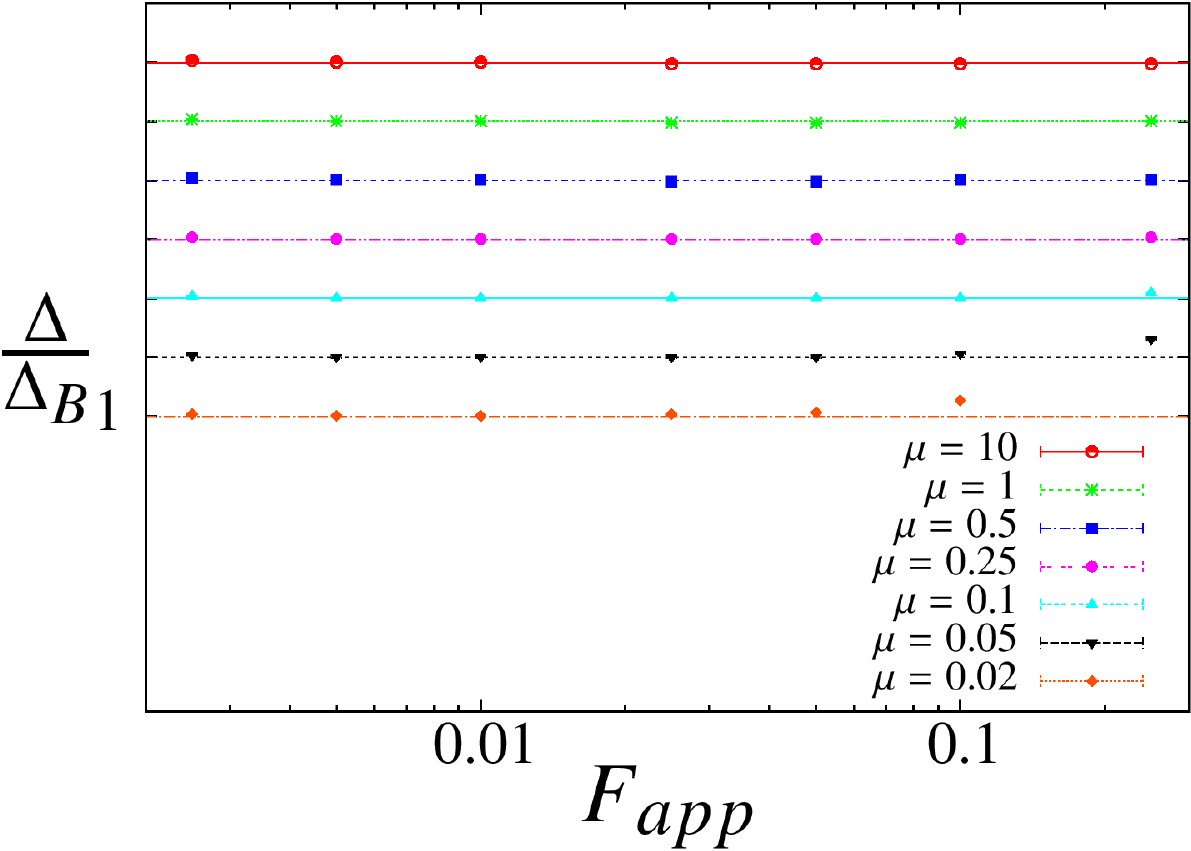}
    \caption{Full width at half height, $\Delta$, of the stress profiles, for
      a range of applied force perturbations and friction coefficients, when a
      single-peak stress response is observed. $\Delta_{B} = 1.113 z$ is the
      theoretical result. For each friction coefficient, data is shifted by
      one unit for convenience. ($\mu=10$ at top descending in order to
      $\mu=0.02$ at bottom.) For $\mu=0.05$ and $F_{app}=0.25 kd$ and
      $\mu=0.02$ and $F_{app}=0.1 kd$, larger full widths of the profiles
      indicating a deviation from the theory. For $\mu=0.02$ and $F_{app}=0.25
      kd$, the profile is shown in Figure~\ref{sfig2} and appears as a
      twin-peak profile.}
    \label{sfig4}
  \end{center}
\end{figure}

To provide a more quantitative comparison between the different plots shown in
Figure \ref{fig1}, as well as to present our results in a manner that is closer
in spirit to similar data obtained from 2d experiments, we show one
dimensional stress profiles in Figures \ref{sfig2}, \ref{sfig3}, and \ref{newfg}.
 These
stress profiles are obtained by circularly averaging the `bottom' stress maps.
Figure \ref{sfig2} shows the stress response profile, $\sigma$, at the bottom
of the box for a relatively large, applied force $F_{\rm app} = 0.25kd$, for
packings with different friction coefficients. These profiles clearly indicate
the role of friction in determining the nature of the stress response. Low
friction results in a twin-peak profile of the anisotropic elastic variety,
that transitions to a single-peak profile at larger friction corresponding to
the isotropic, linear elastic Boussinesq theory \cite{Johnson}. These results
are consistent with similar studies performed in two dimensional geometries
\cite{Nature}.

For a given friction coefficient the stress response is also sensitive to the
magnitude of the perturbing force. We illustrate this point in
Figure~\ref{sfig3}, where we show the stress profiles at the bottom of a
packing with $\mu=0.01$, for different values of $F_{\rm app}$. Again, the
common theme here is that the response changes character depending on $F_{\rm
  app}$. At smaller forces, there is little or no rearrangement in the
particle contacts, and the underlying force network remains intact leading to
a single peak profile. Whereas, for larger forces, a small fraction of
contacts are broken, thereby causing the stresses to be propagated along
better defined paths, leading to twin peak profiles.  For the case where a
single-peak response is obtained, we check that these profiles are consistent
with isotropic, linear elastic theory, by fitting the profiles to the
Boussinesq equation, Eq.~\ref{seq2} as shown in Figure~\ref{newfg}, and
extract the full width at half height. We note that there are no free
parameters in this fitting procedure as the applied forcing is given as an
input. As shown in Figure~\ref{sfig4}, we display our analysis as the ratio
of the fitting to the expected theoretical result. Thus we confirm that the
systems studied here do behave as an isotropic, elastic continuous medium over
a range of parameters in friction and applied forcing.

To gain better insight into this transitional stress response we show the
stress profiles in the plane parallel to the direction of the initial
perturbation in Figure~\ref{fig2}. These two-dimensional plots were obtained
by spherically averaging the stress response about the location of the
perturbation point. This view more clearly demonstrates how an initial
anisotropic response - characterised by the twin-peak stress profile -
transitions towards one with more traditional isotropic features -
characterised by single lobe response profile - and ultimately how the extent
of the force perturbation is eventually, in some sense, dissipated into the
background stress.  Closer inspection of the stress response, as visualised in
Figure~\ref{fig1}, reveals that anisotropic response features can extend
arbitrarily far into the packing depending on the packing parameters.

\subsection{Response Scaling}
\label{sec2}
\begin{figure}[htbp]
  \begin{center} 
    \includegraphics[scale=0.6]{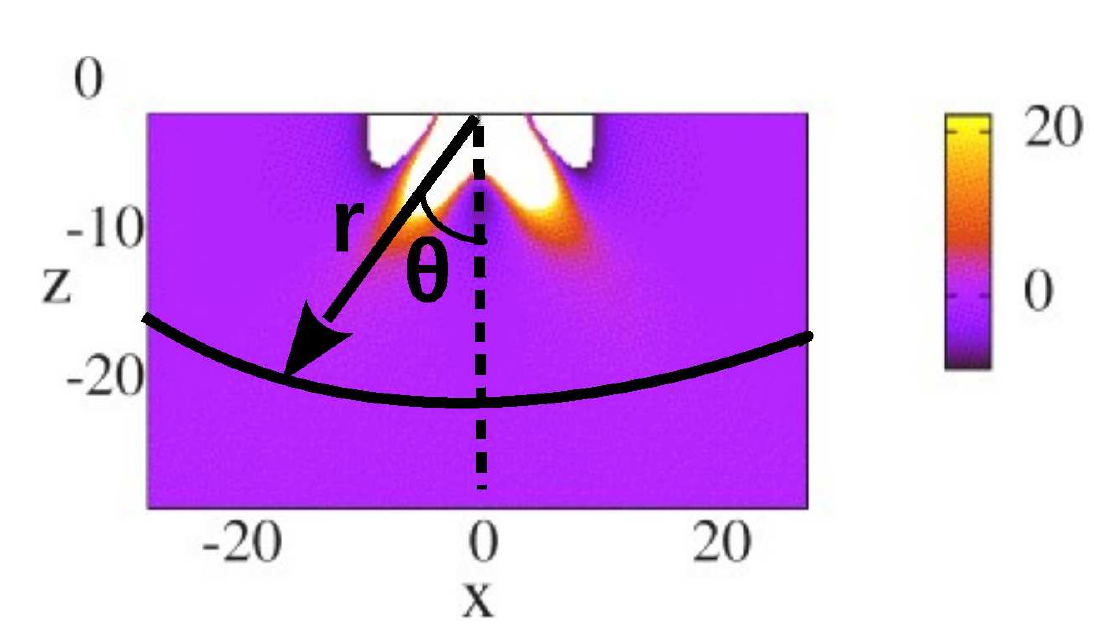}
    \caption{Scanning the circularly average stress response over angle
      $\theta$ at radial slices $r$, to identify the direction of the
      anisotropic response profile relative to the direction for the isotropic
      case (dashed line).}
    \label{sfig5}
  \end{center}
\end{figure}
To characterise the transition between anisotropic and isotropic elastic
stress response behaviours we designed the following procedure that identified
how the maximum stress transmitted through the system deviated from
traditional isotropic behaviour. For an isotropic, linear elastic continuum
described by the Boussinesq equation, the maximum stress occurs in a line
directly ``below'', the point of force perturbation. Our circularly averaged
stress analysis allows us to scan over angle $\theta$, to identify the
direction of maximal stress transmission and monitor this response out to a
radial distance, $r$, from the perturbation source. This construct is shown in
Figure \ref{sfig5}.

\begin{figure}[htbp]
  \begin{center} 
    \includegraphics[scale=0.5]{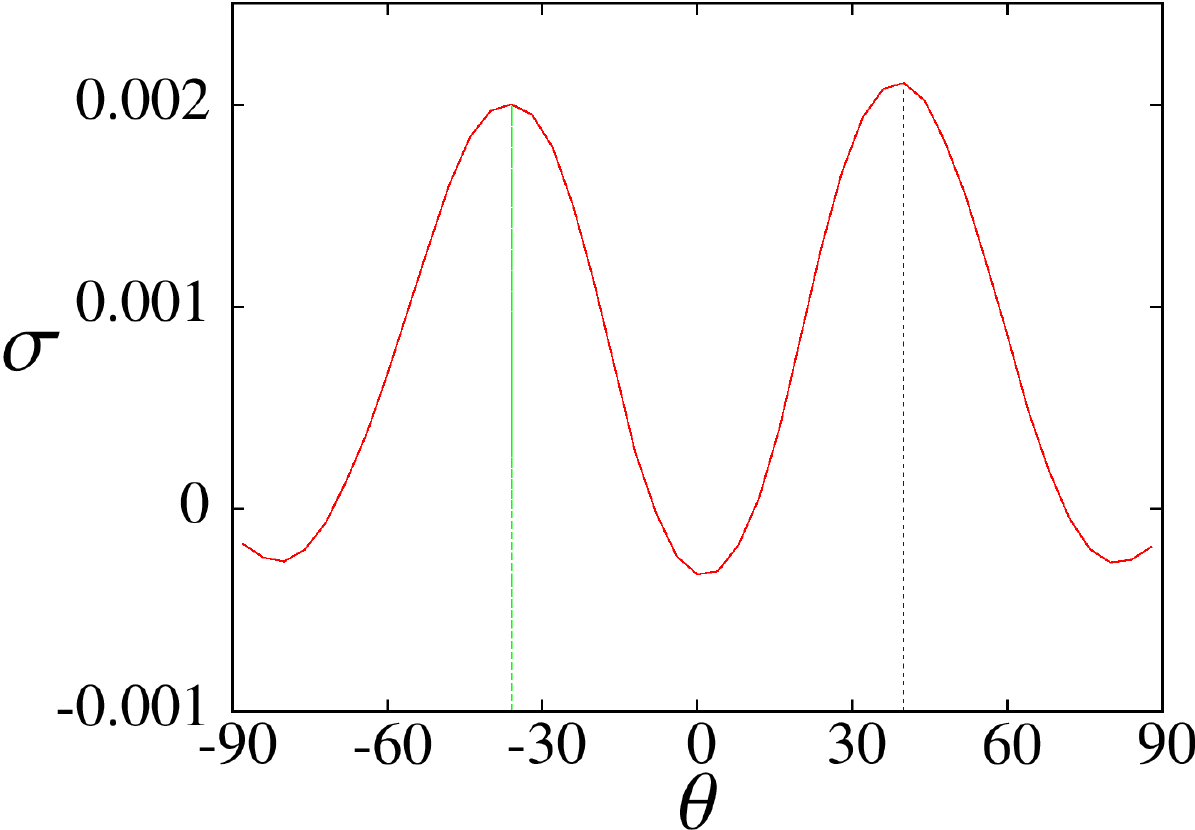}
    \caption{Circularly averaged stress profile, $\sigma$, for a system
      exhibiting an anisotropic, twin-peak profile ($F_{\rm app}=0.25 kd$ and
      $\mu=0.05$). The profile peaks, identified with vertical lines, indicate
      the directions of the maximal stress and are denoted $\theta_{\mu}$
      in the text.}
    \label{sfig6}
  \end{center}
\end{figure}

The anisotropic stress response in our $3D$ geometry transmits through a
cone-like structure. Utilising the azimuthal symmetry of this geometry we
angularly average the stress response, over angular elements $\Delta \theta =
4^{\circ}$, leading to a $2D$ projection of the result as shown in Figure
\ref{sfig6}. The two vertical lines in Figure \ref{sfig6} provides us with a
precise measure of the direction of maximum stress response relative to the
direction for the expected isotropic maximum stress direction ($\theta=0$). We
define the angular direction $\theta_{\mu}$ of maximal stress response as the
average of these two peak locations.
\begin{figure}[htbp]
  \begin{center} 
    \includegraphics[scale=0.5]{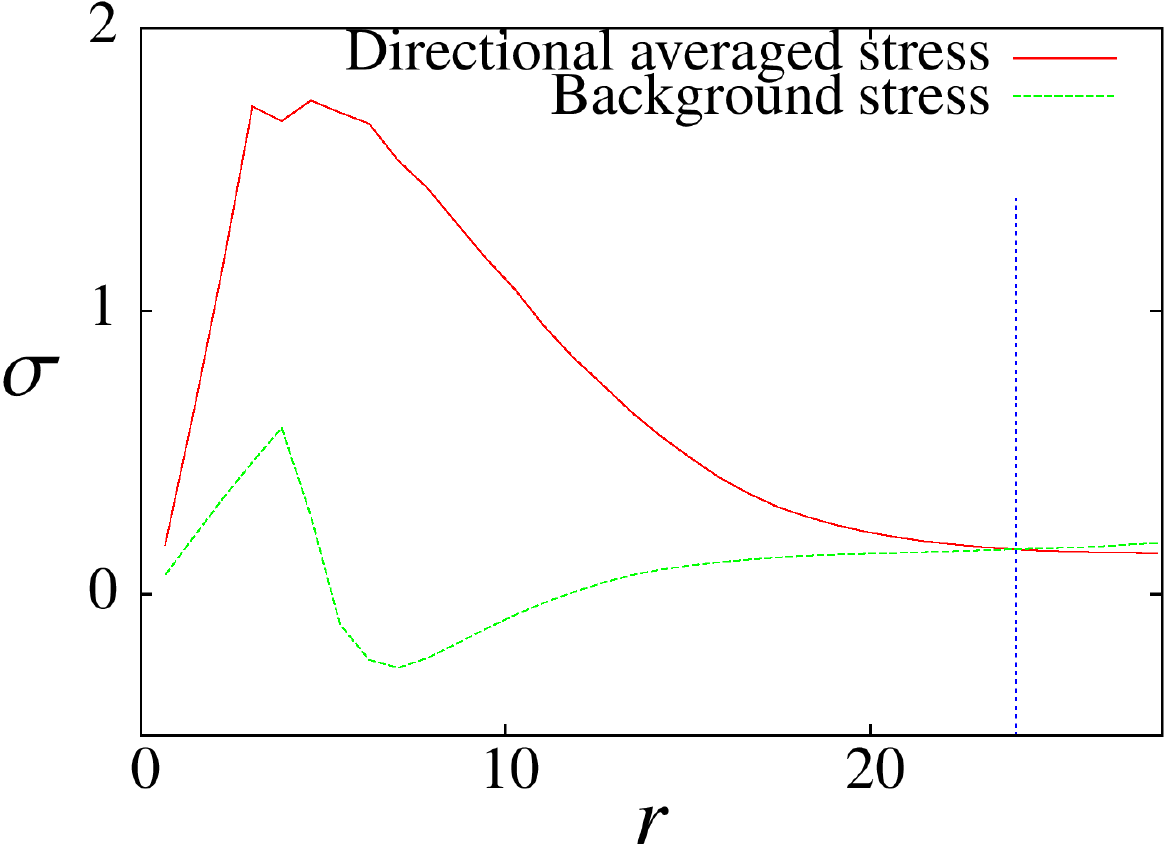}
    \caption{The circularly averaged stress response profile $\sigma$, in the
      direction $\theta_{\mu}$ of maximal stress (red solid line) and the
      background stress (green dashed line) in the direction $\theta=0$,
      identified in Figure \ref{sfig5} (for, $F_{\rm app}=0.25 kd$ and
      $\mu=0.05$). This representative dataset shows how the characteristic
      length scale, $\xi_{\mu}$, at which the directional stress crosses the
      background stress is identified by the vertical line.}
    \label{sfig7}
  \end{center}
\end{figure}
To identify a length scale that quantifies the nature of the mechanical state
we measure the distance, $r$, from the origin of the localised force, in the
direction of maximal stress, until that value of the stress is subsumed into
the background stress. Starting from the angular slice closest to the
perturbation source we scan elemental layers, layer-by-layer, away from the
source to track the direction of the maximum stress. At each layer we compare
the value of the stress in this direction $\theta_{\mu}$ with the stress in
the same radial layer in the direction below the perturbing source,
 $\theta =0$. Thus, we define the characteristic length scale $\xi_{\mu}$ as the
distance at which these two stress values become indistinguishable, as
indicated by the vertical line in Figure \ref{sfig7}.
\begin{figure*}[htbp]
\onecolumngrid
  \begin{center} 
    \includegraphics[scale=0.69]{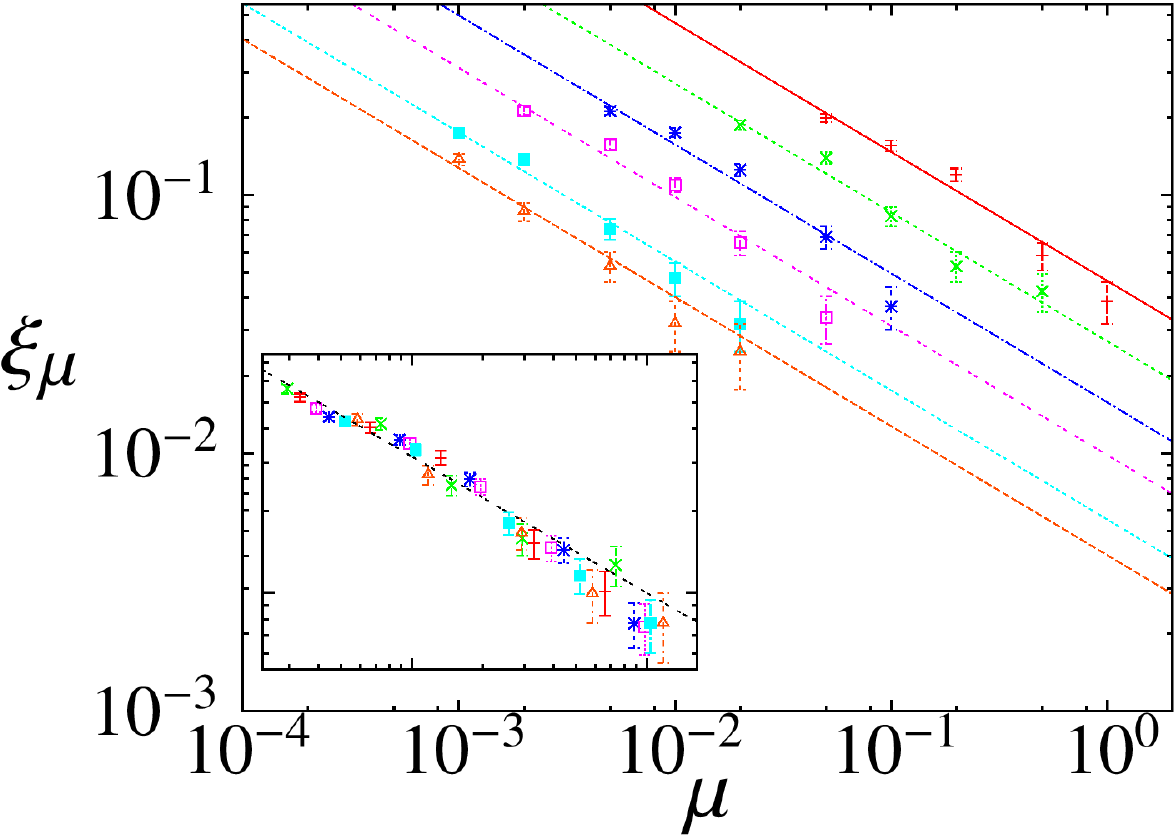}\hfil
    \includegraphics[scale=0.67]{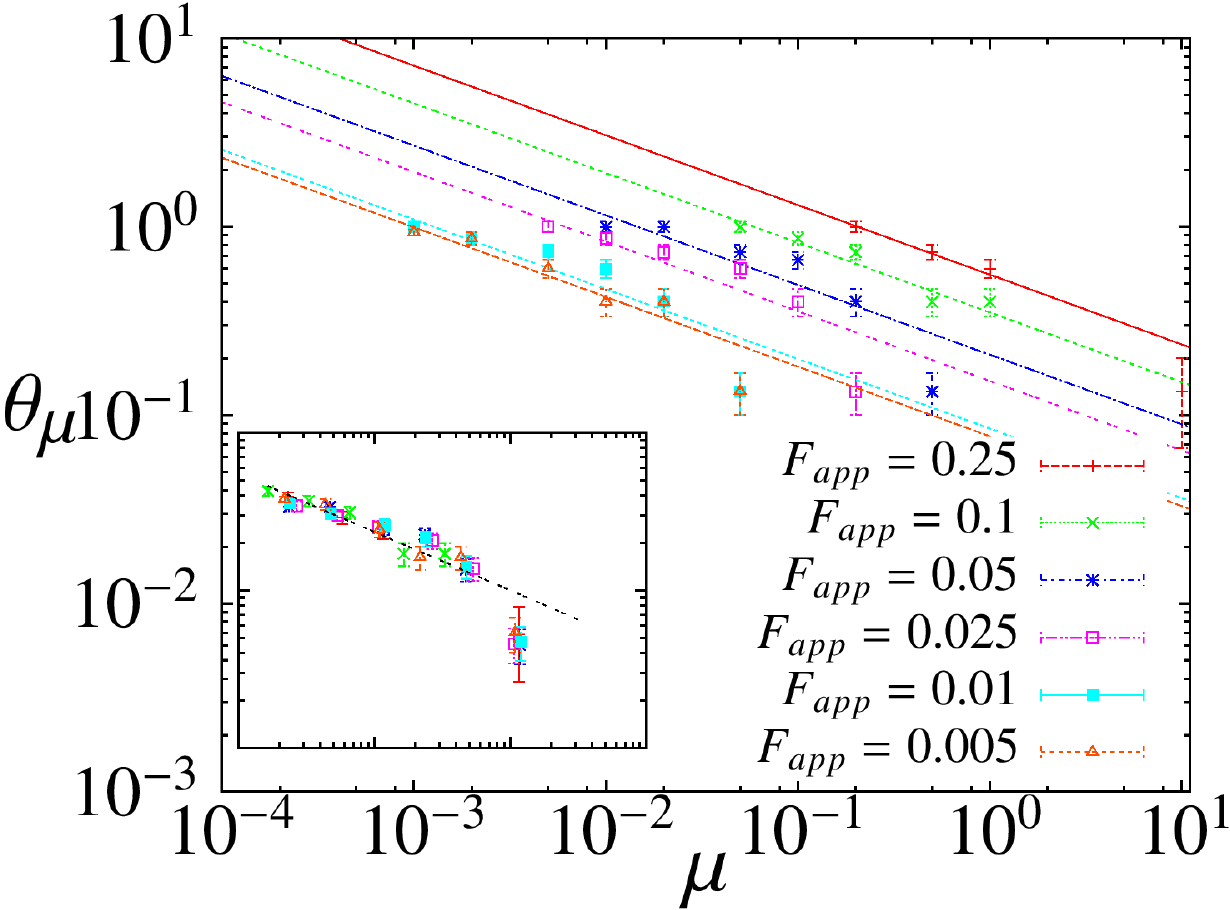}
    \caption{\small Influence of friction and force magnitude on the
      transition between isotropic and anisotropic stress response. (Left)
      Length scale $\xi_{\mu}$ characterising the distance from the location
      of the perturbing force at which the response crosses over from
      anisotropic to isotropic. The lines are power-law fits: $\xi_{\mu} \sim
      \mu^{-\alpha}$, with $\alpha = 0.5 \pm 0.06$. The scaled length data points 
are collapsed onto the line in inset.  (Right) Angle direction of
      the anisotropic stress transmission relative to the central, isotropic
      direction. The lines are power-law fits: $\theta_{\mu} \sim
      \mu^{-\beta}$ for $\mu>0$, with $\beta = 0.37 \pm 0.07$. The scaled angles apart from
 uppermost data points are collapsed onto the line in inset. 
In both panels, the quoted force perturbations are in units of $kd$.}
    \label{fig3}
  \end{center}
\twocolumngrid
\end{figure*}
\begin{figure*}[htbp]
\onecolumngrid
  \begin{center}
    \includegraphics[scale=0.65]{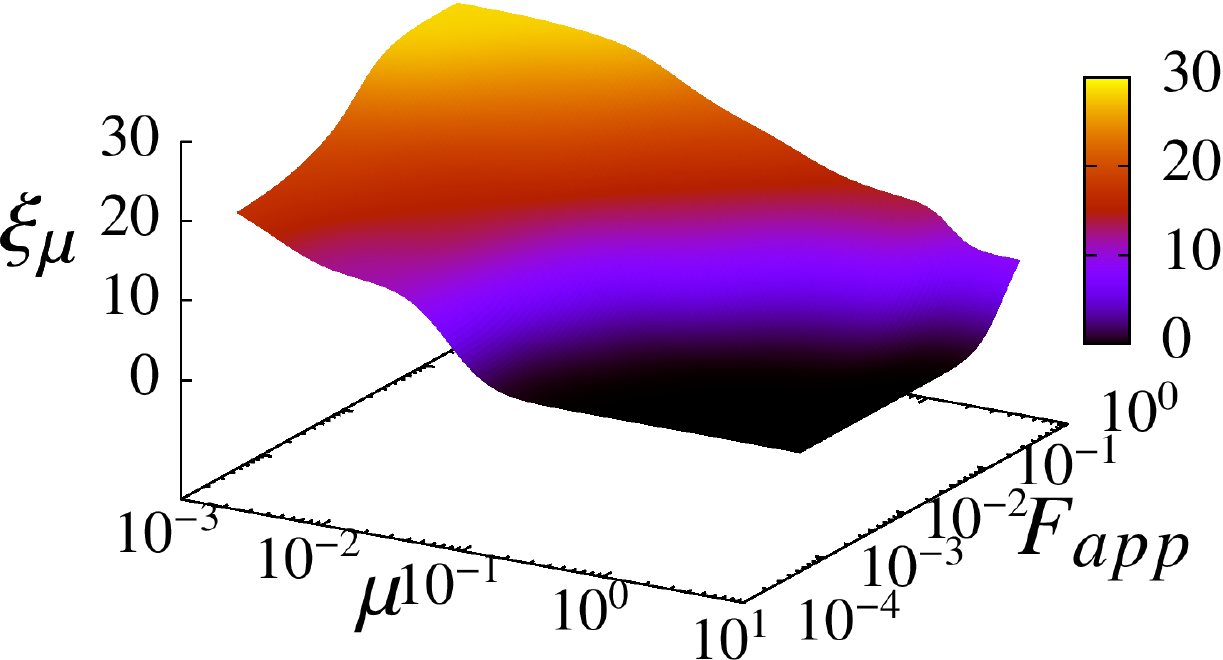}
    \includegraphics[scale=0.65]{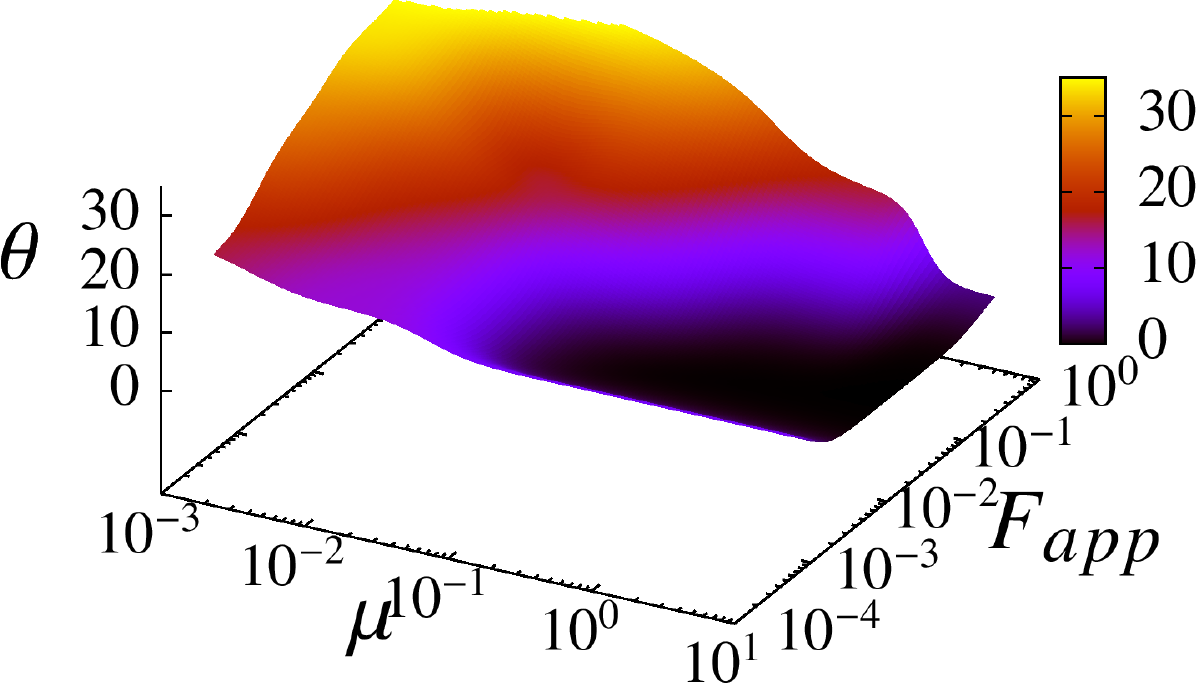}
    \caption{Phase diagrams of force perturbations, in applied force $F_{\rm
        app}$, and friction coefficient $\mu$, for the (a) $\xi_{\mu}$, and (b) stress 
transmission direction $\theta$. These $3D$ plots delineate between regions of
      isotropic, single-peak response (above) and anisotropic, twin-peak,
      response (below the dividing surface).}
    \label{sfig8}
  \end{center}
\twocolumngrid
\end{figure*}

For the extensive parameter space in $\mu$ and $F_{app}$ explored here, we
find that both the characteristic length scale $\xi_{\mu}$, and angle
$\theta_{\mu}$, exhibit power law dependence on friction coefficient of the form
\begin{align}
& \xi_{\mu} \propto \mu^{-\alpha} \\
& \theta_{\mu} \propto \mu^{-\beta} \text{\hspace{0.25in} for $\mu > 0$}
\label{eqn1}
\end{align}
with the power-law exponents, $\alpha=0.5 \pm 0.06$ and $\beta = 0.37 \pm 0.07$,
 over the range of friction and force magnitudes studied.
Thus, the anisotropic-isotropic transition for stress
response of a granular array is characterised by critical-like
features accompanied by a length scale that grows as friction
decreases. A subset of
the data (for visual clarity) is presented in Figure \ref{fig3}. The very
presence of a diverging length scale that distinguishes between different
mechanical phases suggests that the anisotropic-isotropic transition for
stress response of a granular array is characterised by critical-like features
that are controlled by the friction coefficient. Curiously, this behaviour
mimics the manner in which a thermodynamic system behaves in the vicinity of a
continuous phase transition where now the friction coefficient plays the role
of a reduced temperature. In this sense, the length scale defined here
measures the spatial range of fluctuations in the stress over and above the
background isotropic stress. Thus, to some extent the anisotropic-isotropic
transition resembles a one-dimensional Ising model, where the zero-friction
limit corresponds to the ordered phase and infinite friction is the analogue
of the disordered phase, stable fixed point. For any friction `perturbations',
the response is readily driven away from the zero-friction, anisotropic phase,
and always ends up exhibiting isotropic stress response for large enough
system size. 
\begin{figure}[htbp]
  \begin{center} 
    \includegraphics[scale=0.6]{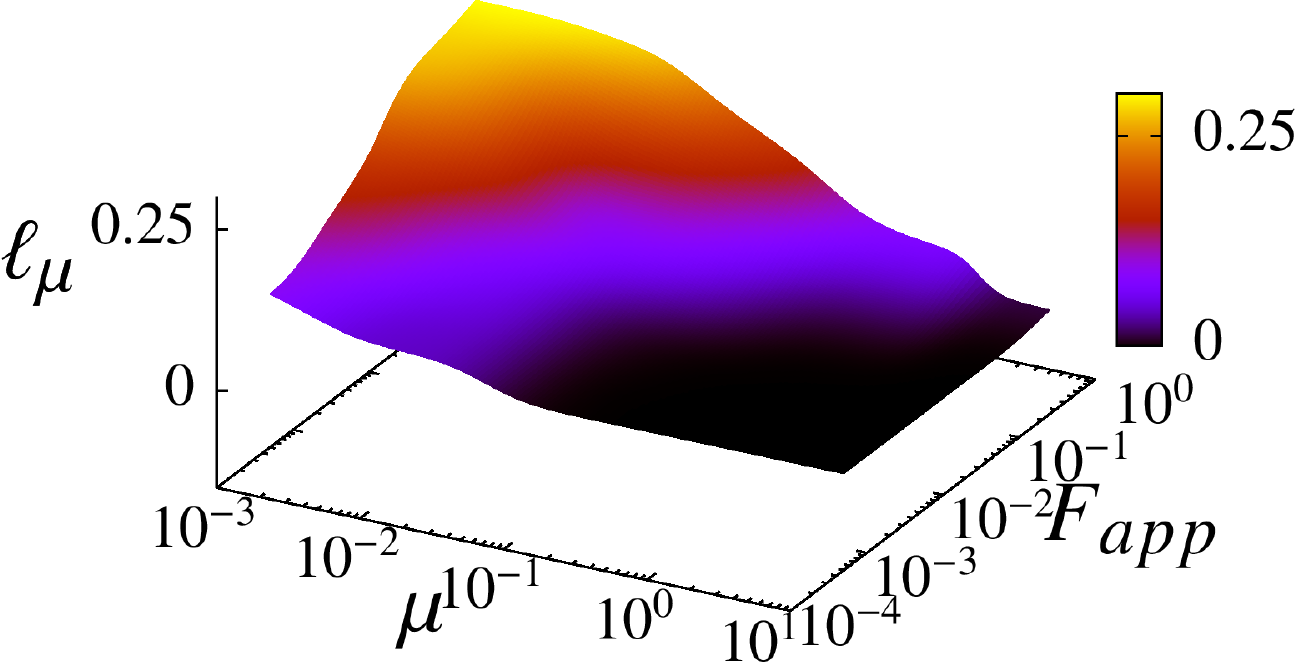}
    \caption{\small Three dimensional phase diagram, in the space of applied
      force $F_{app}$ and particle friction coefficient $\mu$, for the
      anisotropic-isotropic transition of mechanical response within a
      granular array, parameterised by the effective length scale, $\ell_{\mu}
      \equiv \xi_{\mu} \sin \theta$ where $\xi_{\mu}$ is normalised by
      simulation box size.}
    \label{fig4}
  \end{center}
\end{figure}

Upon collating the data over our entire parameter space: $\mu=\{0,10.0\}$ and
$F_{\rm app} =\{0.0005,0.25\}$, we are able to map the stress response
behaviour onto the pseudo-phase diagrams shown in Figure \ref{sfig8}. These
diagrams delineate regions of parameter space where an anisotropic response is
expected (region below the boundary) and those parameters which are more
likely to result in an isotropic, linear elastic response (above boundary).
Anisotropic behaviour emerges for low friction and high forcing where the
stress peaks extend far into the packing and are directed away from
perturbation direction. It is only in the $\mu=0$ limit that the anisotropic
length scale spans the simulation domain. Conversely, isotropic character is
recovered for mildly frictional materials and moderate forcing.

Combining these two plots we present the anisotropic-isotropic transition in
stress response as parameterised by the effective length scale, $\ell_{\mu}
\equiv \xi_{\mu} \sin \theta$, where $\xi_{\mu}$ is normalised by the
      simulation box size, in Figure~\ref{fig4}.
The boundary distinguishes between the stress state within a single
system. For points below the boundary, the granular medium behaves as an
anisotropic, elastic material, where stresses are transmitted through the
system along directed pathways commensurate with the structure of the packing.
Above the boundary, the material will resort to behaviour more consistent with
an isotropic, elastic solid. Thus, small systems subject to relatively large
forces will appear to behave as anisotropic materials that may result in
failure along the directions of the largest stress propagation. Whereas large
frictional aggregates can be considered, as is well-known from everyday
experience tells us, as isotropic elastic media.

\section{Conclusions}
\label{concl}
Large-scale computer simulations of overcompressed FCC granular arrays over a
wider range in friction coefficient and point force perturbations were carried
out to study the stress response inside these arrays. Our results indicate
that packings with large particle friction subject to small localised
forcings, exhibit a single lobe stress response consistent with classical,
isotropic elasticity. Whereas, packings with low friction and subject to large
localised force magnitudes exhibit ring-like stress responses, a hallmark of
anisotropic elastic behaviour. However, this anisotropic state diminishes with
distance from the perturbation source, within the same packing, and
transitions towards a more elastic-like character at a length scale that
depends on the friction coefficient. This characteristic length scale grows
with decreasing friction. Ultimately, the limit of purely frictionless
packings appear to present as a pathological case in this problem - showing
anisotropic behaviour out to the largest distances. Any non-zero friction will
ultimately send the system towards an isotropic elastic regime in the
thermodynamic limit.

While the FCC arrays used in this study clearly enhance the possibility of
observing strongly anisotropic behaviour, the presence of disorder is likely
to suppress such features
\cite{Nature,10.1007/s10035-009-0156-0,10.1007/s10035-009-0156-0}. Thus, we
expect that a combination of varying structural disorder and friction provide
the necessary ingredients for packings that can be constructed with tailored
mechanical properties. It might also be that as a disordered granular system
is brought closer to its point of marginal rigidity, it may become
increasingly fragile
\cite{10.1103/PhysRevLett.81.1841,10.1103/PhysRevE.70.051309}, and remain as
an anisotropic medium at all friction values, thus providing a broad range of
possible mechanical states to choose from.

\acknowledgements 
This work was supported by the National Science Foundation CBET-0828359.

\end{document}